\documentstyle[aps,prb,preprint]{revtex}
\begin{document}
\draft
\tighten

\title{ 
Termination of the spin-resolved integer quantum Hall effect
 }
\author{
  L. W. Wong\cite{email} and H. W. Jiang\cite{email2}
  }
\address{
  Department of Physics,
 University of California at Los Angeles, Los Angeles, Ca 90024
  }
\author{
E. Palm
}
\address{
National High Magnetic Field Laboratory, Tallahassee, Florida 32306
}
\author{
W. J. Schaff
}
\address{
Department of Electrical Engineering, Cornell University, Ithaca, NY 14853
}
\date{\today}

\maketitle

\begin{abstract}
We report a magnetotransport study of the termination of the spin-resolved integer quantum Hall effect by controlled disorder in a gated $GaAs/Al_{x}GaAs_{1-x}$ heterostructure. We have found that, for a given $N^{th}$ Landau level, the difference in filling factors of a pair of spin-split resistivity peaks $\delta\nu_{N} = |\nu_{N\uparrow}$ - $\nu_{N\downarrow}|$ changes rapidly from one to zero near a critical density $n_{c}$.  Scaling analysis shows that  $\delta\nu_{N}$ collapses onto a single curve independent of $N$ when plotted against the parameter $(n-n_{c})/n_{c}$ for five Landau levels.  The effect of increasing the Zeeman energy is also examined by tilting the direction of magnetic field relative to the plane of the two-dimensional electron gas.  Our experiment suggests that the termination of the spin-resolved quantum Hall effect is a phase transition.
\end{abstract}

\pacs{
PACS number(s):
}

In the integer quantum Hall effect (IQHE), when spin of the two-dimensional electron gas (2DEG) is taken into account, each Landau level splits into two levels (up and down spin polarizations) with energy  \(E_{N\uparrow,\downarrow} = (N + 1/2)\hbar\omega_{c}\pm E_{s}\) in the zero disorder limit, where N is the Landau-level index, $E_{s}$ is the spin-split energy.  Contributions to the spin-split energy come from both the exchange interaction and the Zeeman energy.  It is well known from early calculations\cite{s1} and experiments\cite{s2} that $E_{s}$ is dominated by the exchange interaction when the Fermi level lies between spin-split Landau levels.  Thus, in contrast to the IQHE due to the Landau level quantization, the IQHE arises from the spin-split is consequently a many-body effect.
	
Experimentally, it is a well established fact that the spin-resolved  IQHE disappears rapidly in decreasing magnetic field or increasing disorder.  Despite considerable work in the literature relating to spin effects in the IQHE, it remains unclear whether this type of termination is a phase transition.  The interest is renewed recently when the effect of spin on the quantum Hall liquid- insulator transition has became a subject of  debate\cite{s3,s4,s5,s6,s7}.  It was observed experimentally in the $GaAs$ based systems that there is a direct phase transition from an insulator with $\sigma_{xy} = 0$ to a $\nu = 2$ quantum Hall liquid with $\sigma_{xy} = 2 e^{2}/h$ (will be called the 0-2 transition)\cite{s3,s4,s5}.  One aspect of the observation is inconsistent with the widely discussed theoretical global phase diagram proposed by Kivelson, Lee, and Zheng\cite{s8}.  Under the selection rules of the phase diagram, the transition from one IQHE to another or to insulator must change in $\sigma_{xy}$ by a step of $e^{2}/h$ as the delocalized states are assumed to float above the Fermi level one at a time.  Experimentally, there is however no indication of a $\nu = 1$ IQHE regime to separate the 0-2 transition.  In fact, a recent experiment done by Glozman, Johnson and Jiang\cite{s9}, on a low-mobility heterostructure, showed that the $\nu = 1$ IQHE collapses rapidly at an apparent triple point when the disorder of the system is increased continuously.  

The question of whether or not there is a phase transition from spin-resolved IQHE to a spin-degenerate IQHE is crucial in establishing the topology of the phase diagram.  To answer this question , Folger and Shklovskii\cite{s10} have recently proposed a model which suggests that the termination of the spin-resolved IQHE is a second-order phase transition in the absence of Zeeman energy, much like a ferromagnetic to paramagnetic transition.  They argue that the exchange interaction collapses to zero when equal populations of the two spin states are reached due to the disorder-induced level broadening.  The same idea was also discussed by MacDonald and Yang earlier\cite{s11}.  Alternatively, Tikofsky and Kivelson\cite{s12} have proposed a different model that at low n and small B, up and down spin electrons in a even-integer filling factors pair together to form a bosonic state.  This many-body state therefore allows direct phase transitions such as 0-2.  
	
In light of these recently proposed theoretical models, we have conducted a sequence of magneto-transport experiment to study specifically how the spin-resolved IQHE is terminated in the presence of controlled disorder.  We strive to answer the following questions: (1) Is the termination in general a phase transition? (2) What is the topology of the phase diagram in the disorder-magnetic field plane\cite{s9} incorporating with the spin degree of freedom of 2DEG?	
	
Three samples with different mobility used in the experiment were modulation-doped $GaAs/Al_{x}GaAs_{1-x}$ heterostructures fabricated by molecular beam epitaxy. The parameters for these samples are listed in the Table 1.  Hall bar patterns with 3.5 : 1 ratio were etched by standard lithographic techniques and NiCr gates were evaporated onto their surfaces.  The magneto-transport measurements were carried out by standard low-frequency lock-in techniques in a \(^{3}He+^{4}He\) dilution refrigerator with a tilting probe.  With this probe, the angle $\theta$ between the normal of the 2DEG plane with respect to the magnetic field can be varied continuously from 0 to 90$^{\circ}$.  Since the qualitative results of all three samples are identical, we only present here the data from a single sample A for clarity.
	
In the limit of zero-disorder, there exists a peak in the longitudinal conductivity $\sigma_{xx}$ , or equivalently in the longitudinal resistivity $\rho_{xx}$  when the Fermi level concurs with a Landau level.  Thus, one can locate the position of the Landau levels by tracking the peaks in $\rho_{xx}$ when magnetic field B or the density n of the 2DEG is changed.  In the case the spin splits of the $N$-Landau level are well-resolved, peaks corresponding to the spin up and down polarizations in $\rho_{xx}$ are positioned at half-integer values of the average filling factor $\nu_{N} = (2N + 1) \pm 1/2$, or identically $\delta\nu_{N} = |\nu_{N\uparrow}$ - $\nu_{N\downarrow}| = 1$. 
	
The bulk of our experiment was done at the base temperature of 50 mK.  In the course of the experiment, magnetic field B was swept in a very slow rate in one direction to locate peaks of the various spin-split Landau levels in $\rho_{xx}$ while the density n was controlled by a gate voltage.  Their corresponding filling factors are determined by using the values of the magnetic field at the peaks and the density n calculated from the low-field Shubnikov-de Haas oscillations.  The advantage of using high mobility samples in this experiment is that spin-split peaks are well-resolved for a numbered of $N's$ at zero gate voltage.  Fig. 1 shows the typical behavior of the spin up and down peaks of  $N =$ 1, 2, 3, 4, and 5 for different n and B.  At the high density end,  spin-split peaks are well-resolved and $\delta\nu_{N}\approx1$.  $\delta\nu_{N}$ decreases as the density n is reduced and finally, it approaches to zero when the spin up and down peaks merge with each other.  Therefore, it is apparent that the physical parameter $\delta\nu_{N}$  can be used to quantitatively described the separation of a pair spin-split peaks in $\rho_{xx}$.  Experimentally, it is known that by lowering the density n, the Thomas-Fermi screening effect of electrons can be reduced so that the effective disorder in the sample is increased.  By controlling the gate voltage on the samples, we can study the behavior of  $\delta\nu_{N}$ as a function of the density n (or disorder) continuously\cite{s10,s13}.
	
As an example, Fig. 2 shows  the development of  $\delta\nu_{2}$ as a function of the density n.  $\delta\nu_{2}$  remains approximately constant for a broad range of density, and drops from 0.9 to 0.28 abruptly. The data displayed in Fig. 2 are only for those distinctly visible peaks.  The inability of resolving peaks for $\delta\nu_{2}\leq 0.28$ experimentally is believed due to the finite widths of the peaks at finite temperature.  It is very informative to examine at what density $\delta\nu_{2}$  goes to zero.  We have attempted to look for a single empirical function which would give the best fit to all data mathematically.  We have found such a function which gives excellent fitting to the data.  This function resembles the behavior of the Brillouin function, which normally describes the magnetization of a paramagnet in the presence of an external magnetic field\cite{s14}, and it has the following form:
\begin{equation}
\delta\nu_{N}=a\coth[a(n-c)^{1\over2}]+b\coth[b(n-c)^{1\over2}]
\end{equation}		
where a, b, and c are fitting parameters for some $N$. The best fitting parameters are determined by minimizing the $\chi^{2}$ value with tolerance value of 0.01 using the Levenberg-Marquardt algorithm\cite{s15}.  It is worth of noting that   \(|a-b| = 1\pm0.10\) for all $N's$ for our fittings. The density $n_{c}$ at which $\delta\nu_{N}$  equals to zero is evaluated using the equation with the best fitting parameters. The error bar of $n_{c}$ is determined by using the combinations of the standard errors of the fitting parameters.  From the fitting, there appears to be  a critical density $n_{c}$ such that $\delta\nu_{2}$ equals to zero for $n\leq n_{c}$.  $\delta\nu_{N}$ behaves similarly for $N =$ 1, 3, 4 and 5.
	
It is important at this point to mention that the possibility to observe a pair of distinct peaks for the  $N$-Landau level depends not only on the separation of the peaks, but also on the temperature at which the measurement is performed.  As temperature is increased, both the  height and width of the peaks rise rapidly\cite{s16}.  In order to examine whether the finite temperature effect obscures our ability to observe the critical behavior and the existence of the critical density $n_{c}$, we measure $\delta\nu_{2}$ as a function of the density n at three different temperatures. The inset shows two pieces of evidence in favor of a phase transition in the limit of  $T\rightarrow0$ : (1) There exists a common critical density $n_{c}$ within the error bar of the fitting($\delta\nu_{2} = 0$ for $n \leq n_{c}$) for all three temperatures, (2) As $T$ reduces, $\delta\nu_{2}$ approaches to zero more rapidly as $n\rightarrow n_{c}$.

We have performed a scaling analysis to examine the seemingly critical behavior of the data for all spin-resolved Landau levels of $N$ = 1, 2, 3, 4, and 5.  We found that the data collapse onto a single curve when $\delta\nu_{N}$ is plotted against the dimensionless parameter $(n-n_{c})/n_{c}$.  In this exercise, similar to the case for $N = 2$, $n_{c}$$'s$ are determined by extrapolating $\delta\nu_{N}$ to zero using the above method for all other $N's$.  The finding that the curves from different Landau levels can be scaled to a single curve at same temperature suggest that the relevant energy scales for the spin-resolved IQHE are the same for different Landau levels near their respective critical densities.  An approximately linear relation between $N$ and $n_{c}$ is shown in the insert of Fig. 3.  The linearity represents an interesting experimental fact that the critical magnetic-field $B_{c}$ is the same for different Landau levels for a given sample. The values of $B_{c}$ for all three samples are listed in Table 1.  It is worth of noting that a $N\propto n^{5/6}$ dependence has been predicted for heterostructures at moderate densities\cite{s10}.
	
Our findings suggest, on the experimental ground, that the termination of the spin-resolved IQHE is a phase transition.  With this assumption, a phase diagram in the disorder-magnetic field can be constructed as shown in Fig. 4.  Because the ability of the 2DEG to screen the random potential depends inversely on density, the inverse density can also be used as a qualitative measure of disorder.  It is be noticed from Fig. 4 that all the spin-split Landau levels behave the same in the disorder-magnetic plane.
	
Next, we would like to examine the effect of Zeeman energy.  The Zeeman energy is expected to play a critical role in the termination of the spin-resolved IQHE\cite{s10,s12}.   In the model proposed by Fogler and Shklovskii\cite{s10}, high disorder induces the overlapping of the impurity broadened spin-split $N$-Landau levels causing the destruction of the exchange energy \(E_{ex}\propto\Delta n_{N} = |n_{N\uparrow}$ - $n_{N\downarrow}|\).  A second-order phase transition from the  ferromagnetic phase to the paramagnetic phase is expected in the absence of Zeeman energy.  However, the phase transition is expected to be smeared out (i.e. $\delta\nu_{N}$ curve should be tailed-off).  In another model proposed by Tikofsky and Kivelson\cite{s12}, there is a IQHE at even-integers originated from electron-electron interactions.  They argue that, at low density $n$ and magnetic field $B$, it is energetically more favorable for electrons with up and down spin together to form bosons that can condense into bosonic many-body IQHE.  In this model, the stability of the spin-split Landau levels relies on that $E_{s}$ works as a pair breaker preventing the up and down spin electrons from forming pairs at high magnetic field $B$.  When $E_{s}$ is less than some spin gap $\Delta_{s}$, paired states are energetic favorable.  Thus, they predict that the increasing $E_{s}$ will shift the critical density or magnetic field towards lower value.
	
Experimentally, the Zeeman energy is always finite in the $GaAs/Al_{x}GaAs_{1-x}$ heterostructures.  However, as mentioned in the introduction, the spin-split energy $E_{s}$ is dominated by the exchange interaction.  $E_{s}$ can be written as \(g_{0}\mu_{B}B + E_{ex}\), where $g_{0}\mu_{B}B$ is the Zeeman energy and $E_{ex}$ is the exchange energy, and $E_{ex}$ is proportional to the difference of the occupations of the two spin-split $N$-Landau levels $\Delta n_{N} = |n_{N\uparrow}$ - $n_{N\downarrow}|$ at the Fermi level.  Experimentally, the ratio $E_{ex}/g_{0}\mu_{B}B$ is found as large as twenty at odd integer filling factors \cite{s17}.  Since the Zeeman energy $g_{0}\mu_{B}B$ depends on the total magnetic field $B$ while the exchange energy depending only on the magnetic field perpendicular to the plane of the 2DEG, we can increase $g_{0}\mu_{B}B$ by tilting the magnetic field $B$ away from the direction normal to the sample's surface.  In Fig. 5, measurements of $\delta\nu_{2}$ as a function of the density n are shown for five tilting angles $\theta$ = 0, 64.4, 74.4, and 81.9 degrees.  For rough estimation, they correspond to $E_{ex}/g_{0}\mu_{B}B$ approximately of 16, 7.0, 4.5, and 2.5.  The data suggests that the phase transition is robust against the increasing Zeeman energy.  The larger Zeeman energy tends to prevent the termination of the spin-resolved IQHE until lower density is reached.  Unfortunately, the tail-off  behavior of $\delta\nu_{2}$(i.e., the smearing of the phase transition) is not observed in our data for a small exchange to Zeeman energy ratio of two.  However, this is possibly due to our experimental inability to resolve $\delta\nu_{N}\leq0.25$ at our lowest temperature of 50 mK\cite{s18}.  Therefore, whether of not there is a true phase transition in the presence of finite Zeeman energy remains to be determined in the future. 
	
In summary, we have investigated the termination of the spin-resolved IQHE in gated $GaAs/Al_{x}GaAs_{1-x}$ heterostructures.  Scaling behavior is observed when $\delta\nu_{N}$ is plotted against the reduced variable $(n-n_{c})/n_{c}$ for five Landau levels.  By increasing the Zeeman energy, we found the value of $n_{c}$ decreases accordingly.  There is no smearing to the transition when Zeeman energy is increased as many as a factor of eight.  Our data suggest that the termination of the spin-resolved IQHE due to the disorder is a phase transition.  Evidently, the physical origin of the suggested phase transition cannot be determined by the current experiment.  This is left for future studies.
	
We would like to thank M. Fogler, B. Shklovskii S. Kivelson, A. Tikofsky for useful discussions; and D. Chang for helping us in sample preparations.  This work is supported by the NSF under grant number DMR-93-13786; a portion of  this work was performed at the National High Magnetic Field Laboratory, which is supported by NSF Cooperative Agreement No. DMR-9016241 and by the State of Florida.

\begin{table}
\caption{Parameters of the samples used in our experiment.
\label{table1}}
\begin{tabular}{|l|l|l|l|l|}					
sample	&spacer ($nm$)	&mobility ($cm^{2}/vs$)	&density ($10^{11}/cm^{2}$)	&critical field $B_{c}$ ($T$)\\					\hline
A	&50	&0.9$\times10^{6}$	&1.96	&$\approx0.5$\\	\hline
B	&5	&1.2$\times10^{5}$	&7.0	&$\approx3.0$\\	\hline
C	&0	&3.0$\times10^{4}$	&4.5	&$\approx4.5$\\	
\end{tabular}
\end{table}

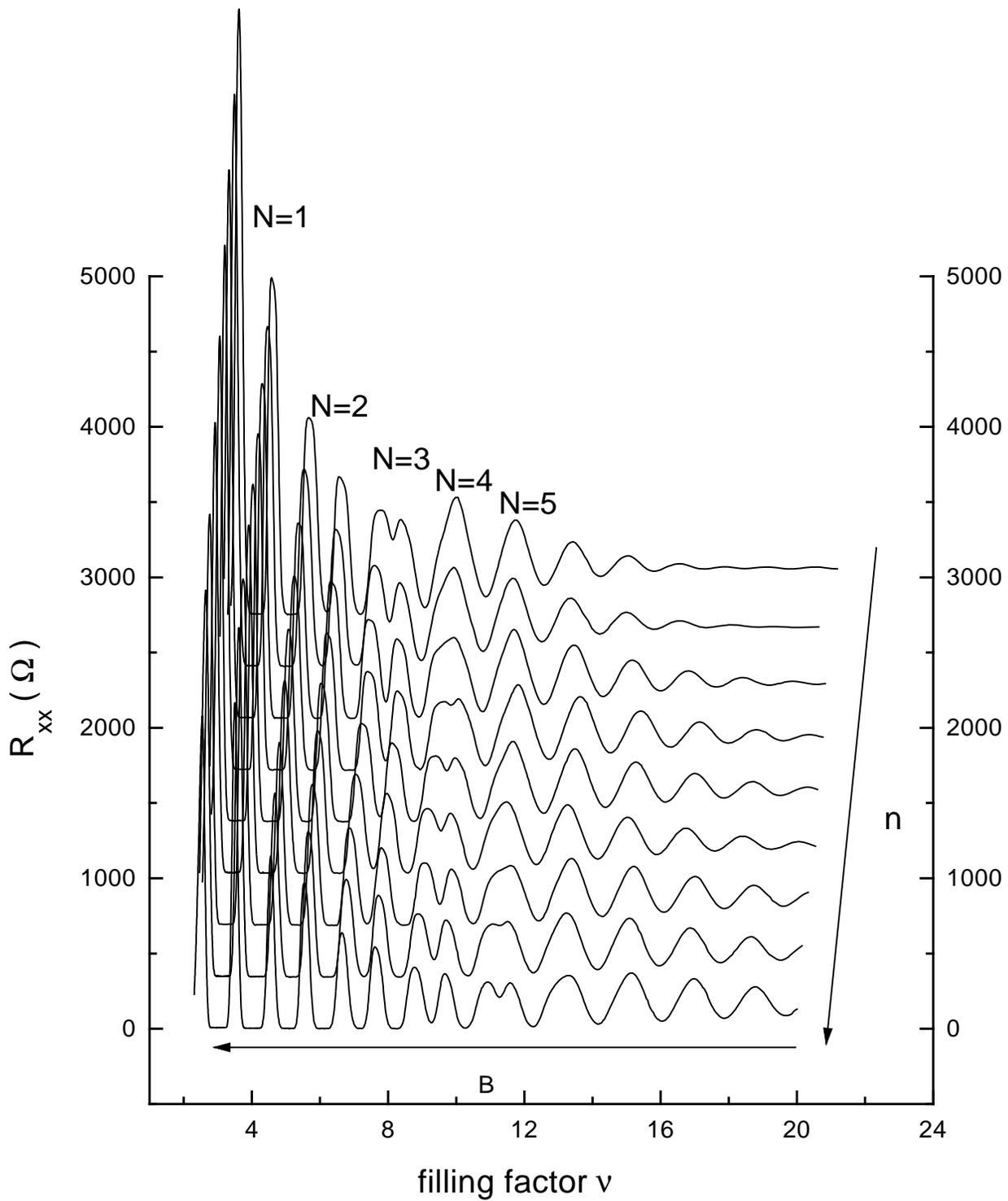
\begin{figure}
\caption{
Pseudo-3D plot of the magnetoresistance curves $R_{xx}$ showing the evolution of the spin-split peaks as a function of the electron density $n$ and magnetic field $B$.
}
\end{figure}

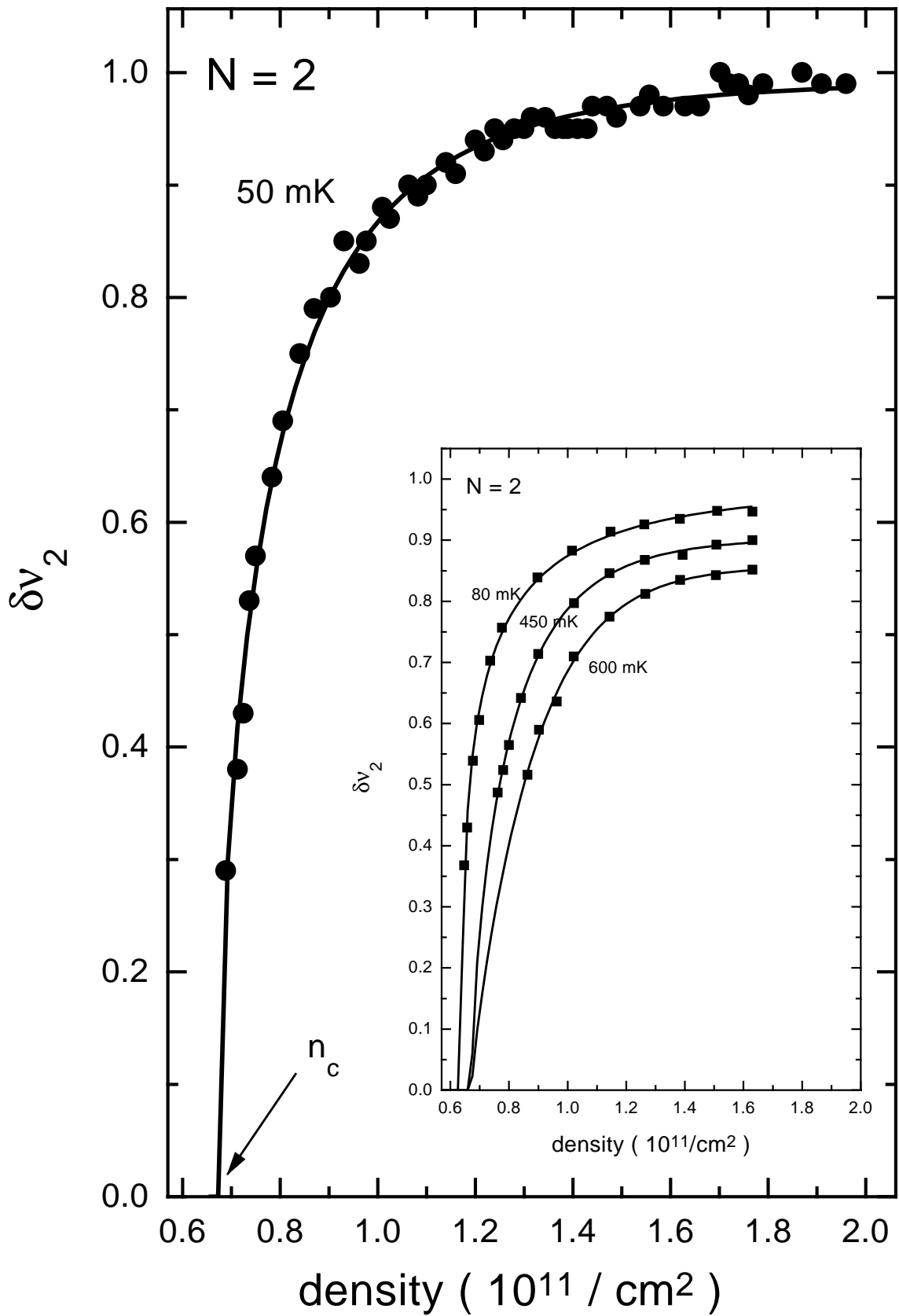
\begin{figure}
\caption{
The quantity $\delta\nu_{2}$ is plotted as a function of the density $n$ at 50 mK. Insert shows $\delta\nu_{2}$ as a function of the density $n$ for three different temperatures: 80, 450, and 600 mK; and the solid lines are the best fitting curves of the form given by Eq. 1.
}
\end{figure}

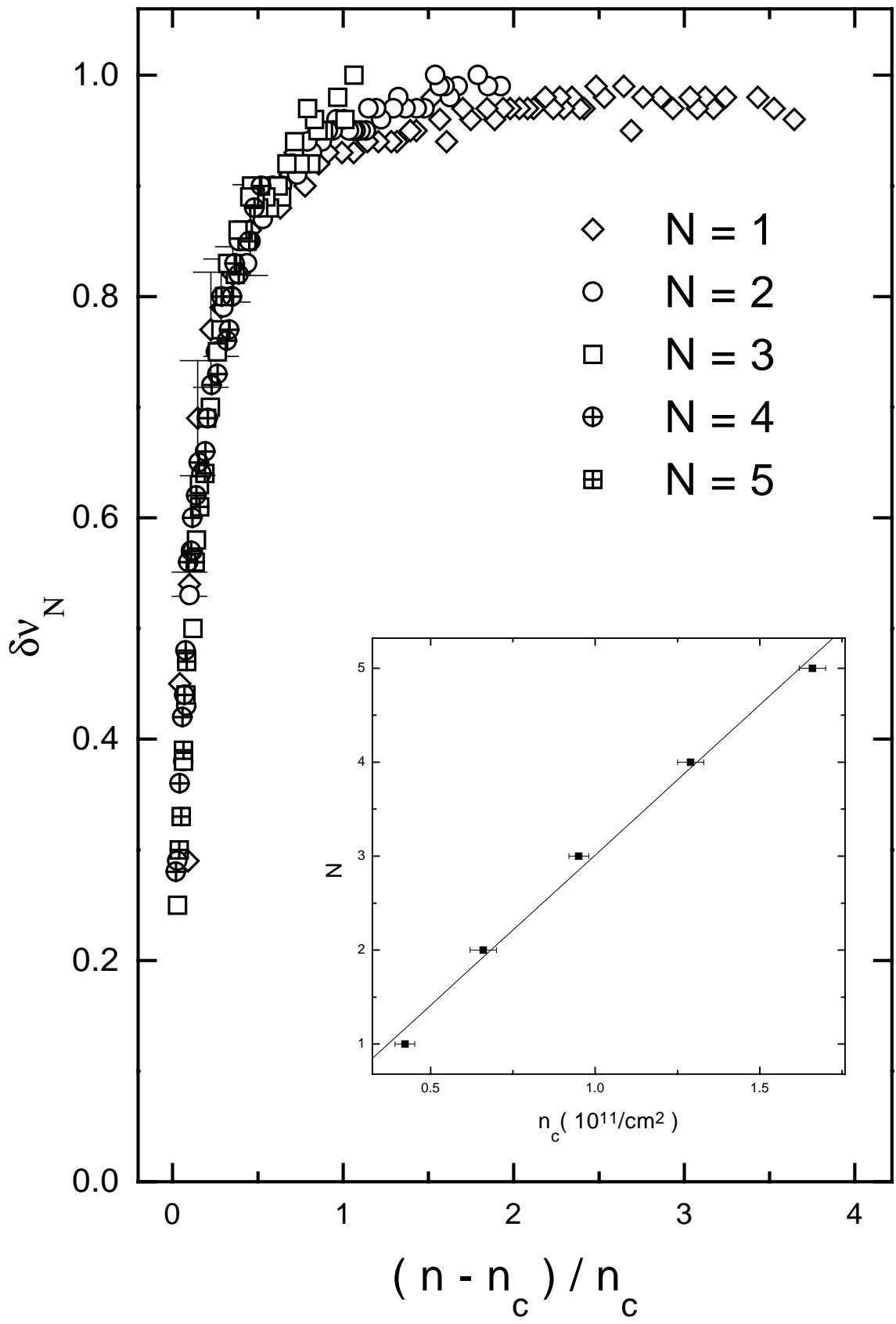
\begin{figure}
\caption{
Scaling analysis shows data collapsing onto a single curve when $\delta\nu_{N}$ is plotted against the dimensionless parameter $(n - n_{c})/n_{c}$ for $N =$ 1, 2, 3, 4, and 5.  Insert shows the linear relation of $N$ and $n_{c}$ with uncertainties given by the error bars.
}
\end{figure}
  
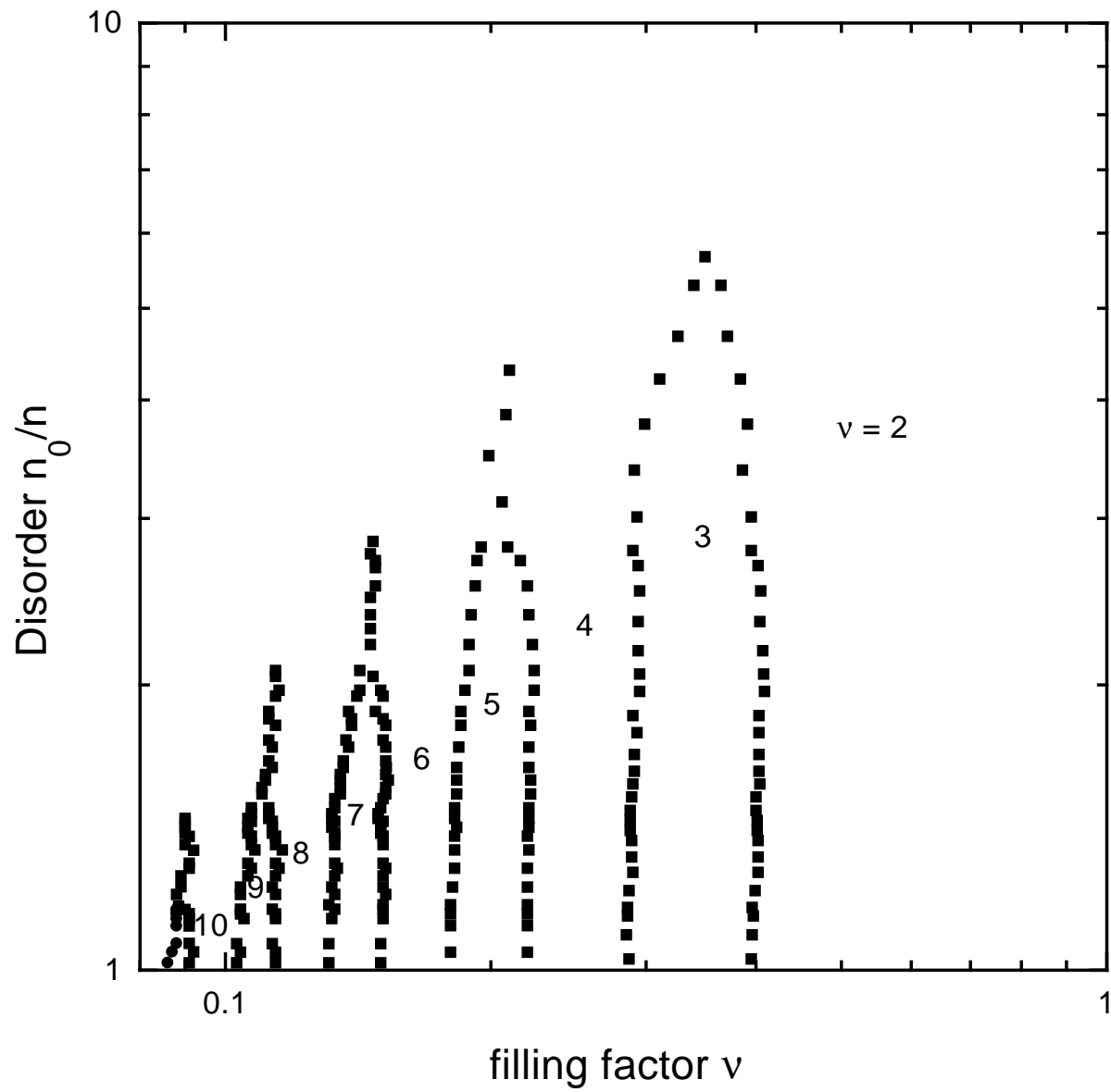
\begin{figure}
\caption{
A portion of the phase diagram in the disorder-magnetic field plane incorporated with the spin degree of freedom in the IQHE. The vertical axis, $n_{0}/n$, where $n_{0} = 1.96\times10^{11}/cm^{2}$, is meant to represent qualitatively the amount of disorder.  Notice that all the spin-split Landau levels behave very similar.
}
\end{figure}

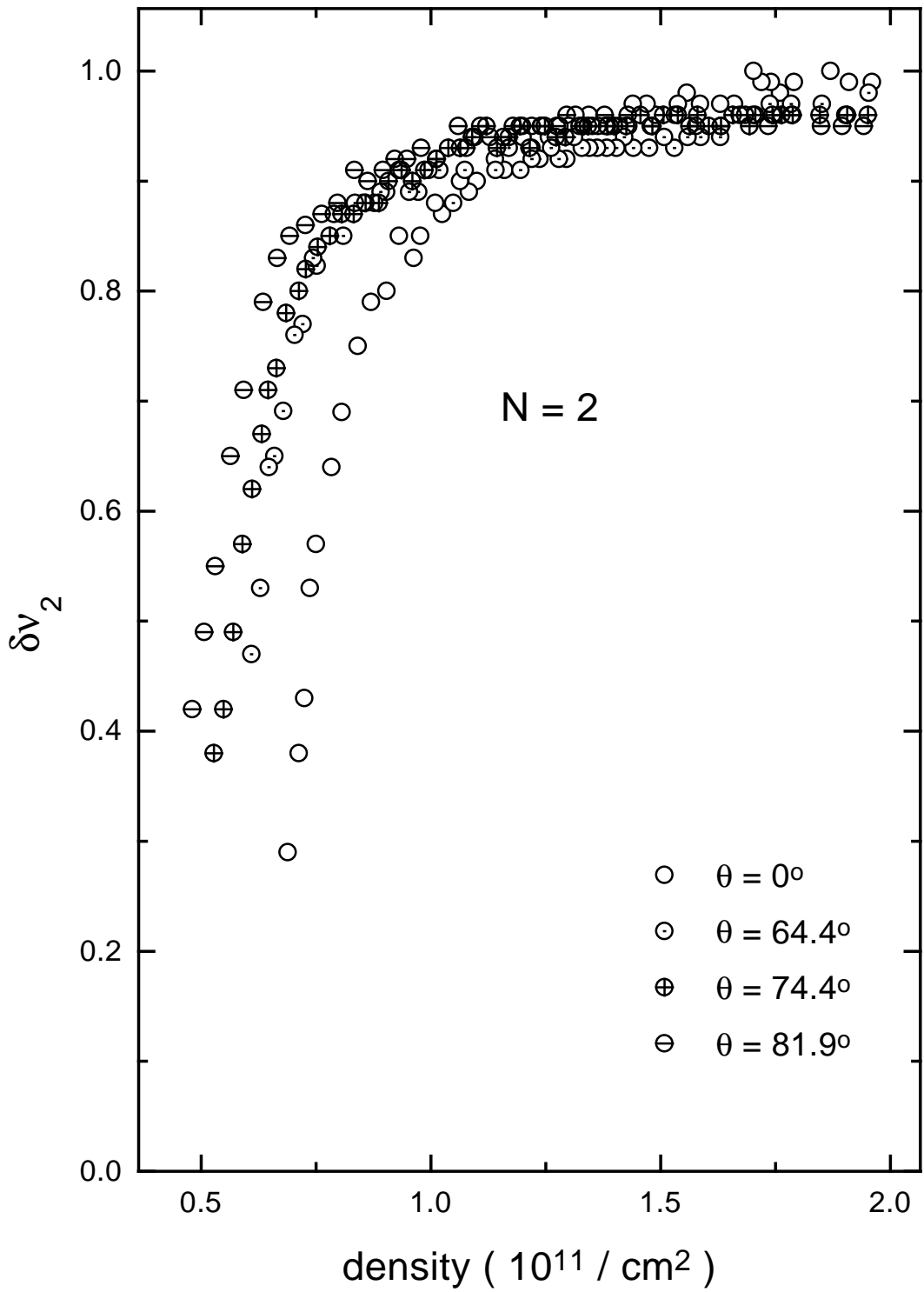
\begin{figure}
\caption{
Effect of increasing the bare Zeeman energy $g_{0}\mu_{B}B$ by tilting method for four different tilting angles $\theta$: 0, 64.4, 74.4, and 81.9 degrees.
}
\end{figure}


\begin{references}
\bibitem{s18}A calculation for the case of our largest tilting angel of 81.9$^{\circ}$ $(E_{ex}/g_{0}\mu_{B}\approx2.5)$ has shown that the tail-off behavior is only significant when $\delta\nu_{N}< 0.2$, M. Fogler, private communication.
\end{references}
\end{document}